\documentclass[aps,pra,twocolumn,superscriptaddress,longbibliography]{revtex4-1}
\usepackage{graphicx}
\usepackage{latexsym}
\usepackage{amssymb}
\usepackage{amsmath}
\usepackage{amsfonts}
\usepackage{upgreek}
\usepackage{subfigure}
\usepackage{bm}
\usepackage{mhchem}
\usepackage{nicefrac}
\usepackage{bbold}
\usepackage{verbatim}
\usepackage[unicode=true,
 bookmarks=false,
 breaklinks=false,pdfborder={0 0 1},backref=false,colorlinks=true]
 {hyperref}
\hypersetup{
 linkcolor=[rgb]{0,0,1},citecolor=[rgb]{0,0,1},urlcolor=[rgb]{0,0,1}}
\usepackage{multirow}
\usepackage{color}
\usepackage{comment}
\usepackage{xcolor}
\usepackage{soul}
\usepackage{tikz}

\def\ciup{\hat{c}_{\mathbf{i}, \uparrow}^{\vphantom\dagger}}
\def\cjup{\hat{c}_{\mathbf{j}, \uparrow}^{\vphantom\dagger}}
\def\cjdup{\hat{c}_{\mathbf{j}, \uparrow}^{\dagger}}
\def\cidup{\hat{c}_{\mathbf{i}, \uparrow}^{\dagger}}

\def\cidown{\hat{c}_{\mathbf{i}, \downarrow}^{\vphantom\dagger}}
\def\cjdown{\hat{c}_{\mathbf{j}, \downarrow}^{\vphantom\dagger}}
\def\cjddown{\hat{c}_{\mathbf{j}, \downarrow}^{\dagger}}
\def\ciddown{\hat{c}_{\mathbf{i}, \downarrow}^{\dagger}}

\def\nidown{\hat{n}_{\mathbf{i}, \downarrow}}
\def\niup{\hat{n}_{\mathbf{i}, \uparrow}}

\def\njdown{\hat{n}_{\mathbf{j}, \downarrow}}
\def\njup{\hat{n}_{\mathbf{j}, \uparrow}}

\def\ni{\hat{n}_{\mathbf{i}}}
\def\nj{\hat{n}_{\mathbf{j}}}

\usepackage{braket}

\begin{document}

\title{Sub-dimensional magnetic polarons in the one-hole doped $\rm{\mathbf{SU(3)}}$ $\mathbf{t}$-$\mathbf{J}$ model}

\author{Henning Schl\"omer}
\affiliation{Department of Physics and Arnold Sommerfeld Center for Theoretical Physics (ASC), Ludwig-Maximilians-Universit\"at M\"unchen, Theresienstr. 37, M\"unchen D-80333, Germany}
\affiliation{Munich Center for Quantum Science and Technology (MCQST), Schellingstr. 4, D-80799 M\"unchen, Germany}
\affiliation{Department of Physics and Astronomy, Rice University, Houston, Texas 77005, USA}
\author{Fabian Grusdt}
\affiliation{Department of Physics and Arnold Sommerfeld Center for Theoretical Physics (ASC), Ludwig-Maximilians-Universit\"at M\"unchen, Theresienstr. 37, M\"unchen D-80333, Germany}
\affiliation{Munich Center for Quantum Science and Technology (MCQST), Schellingstr. 4, D-80799 M\"unchen, Germany}
\author{Ulrich Schollw\"ock}
\affiliation{Department of Physics and Arnold Sommerfeld Center for Theoretical Physics (ASC), Ludwig-Maximilians-Universit\"at M\"unchen, Theresienstr. 37, M\"unchen D-80333, Germany}
\affiliation{Munich Center for Quantum Science and Technology (MCQST), Schellingstr. 4, D-80799 M\"unchen, Germany}
\author{Kaden R. A. Hazzard}
\affiliation{Department of Physics and Astronomy, Rice University, Houston, Texas 77005, USA}
\author{Annabelle Bohrdt}
\affiliation{Munich Center for Quantum Science and Technology (MCQST), Schellingstr. 4, D-80799 M\"unchen, Germany}
\affiliation{University of Regensburg, Universit\"atsstr. 31, Regensburg D-93053, Germany}

\date{\today}
\begin{abstract}
The physics of doped Mott insulators is at the heart of strongly correlated materials and is believed to constitute an essential ingredient for high-temperature superconductivity. In systems with higher $\rm{SU(}$$N)$ spin symmetries, even richer magnetic ground states appear at a filling of one particle per site compared to the case of $\rm{SU(2)}$ spins, but their fate upon doping remains largely unexplored. Here we address this question by studying a single hole in the SU(3) $t$-$J$ model, whose undoped ground state features long-range, diagonal spin stripes. By analyzing both ground state and dynamical properties utilizing the density matrix renormalization group, we establish the appearence of magnetic polarons consisting of chargons and flavor defects, whose dynamics is constrained to a single effective dimension along the ordered diagonal. We semi-analytically describe the system using geometric string theory, where paths of hole motion are the fundamental degrees of freedom. With recent advances in the realization and control of $\rm{SU(}$$N)$ Fermi-Hubbard models with ultracold atoms in optical lattices, our results can directly be observed in quantum gas microscopes with single-site resolution. Our work suggests the appearance of intricate ground states at finite doping constituted by emergent, coupled Luttinger liquids along diagonals, and is a first step towards exploring a wealth of physics in doped $\rm{SU(}$$N)$ Fermi-Hubbard models on various geometries.
\end{abstract}
\maketitle

\section{Introduction} The physics of doped Mott insulators, dominated by the intricate competition between kinetic and interaction energies, is at the heart of strongly correlated physics. For two-flavor spins, the two-dimensional (2D) $\rm{SU(2)}$ symmetric Fermi-Hubbard (FH) model represents a paradigmatic model believed to capture some essential physics of high-temperature superconducting cuprates~\cite{Lee2006, Keimer2015}, and tremendous progress has been made both numerically~\cite{LeBlanc2015, Zheng2017, Jiang2019_science, Qin2020, MultiMess2021, Jiang_Kivelson, Arovas2022_rev} and experimentally~\cite{jordens2008mott, schneider2008metallic, Bloch2008, Bloch2012, Hart2015, Cheuk2015, Parsons2015, Gross2017, Schafer2020}.

Models with higher spin symmetries, realized by $\rm{SU(}$$N>2)$ invariant models, promise rich physics beyond the $\rm{SU(2)}$ paradigm~\cite{Affleck1988, Honerkamp2004, Assaad2005, Hermele2009, Sotnikov2014, Sotnikov2015, Yanatori2016, Torbati2018, Torbati2019, Eduardo_universal}. Indeed, they have established as important generalizations of the SU(2) symmetric FH model to describe systems with, e.g., orbital degeneracy. In particular, $N>2$ models are relevant to capture the essential physics of transition-metal oxides~\cite{kugel1973crystal, Li1998, Tokura2000}, the Kondo effect~\cite{Doniach1977, Coleman1983} and heavy fermion compounds~\cite{Hewson_1993, coleman2015introduction}, as well as single and twisted bilayer graphene~\cite{Goerbig2011, Xu2018tw, Liao2021, Natori2019}.

In addition to novel magnetic states and enhanced quantum fluctuations, SU($N$) symmetric models separate features that are linked in the vanilla $\rm{SU(2)}$ Hubbard model: perfect nesting and van Hove singularities at one particle per site are absent, in contrast to the $\rm{SU(2)}$ FH model. At unit filling $\braket{\hat{n}} = 1$, the $\rm{SU(3)}$ FH model has been shown to feature rich magnetic structures of various translation symmetry breaking patterns, where in particular finite repulsive interactions $U>0$ are necessary to open a charge gap and observe magnetically ordered Mott insulating (MI) states~\cite{Gorelik2009, Sotnikov2014, Sotnikov2015, Chunhan2023, Eduardo2023}. This is in contrast to the SU(2) FH model, where metal-to-MI transitions are absent. In the strongly coupled regime of the SU(3) FH model, effective mappings to $\rm{SU(3)}$ Heisenberg models at $\braket{\hat{n}} = 1$ reveal a 3-sublattice (3-SL), diagonally striped magnetic order, stabilized by quantum fluctuations through an order by disorder mechanism~\cite{Toth2010, Bauer2012}.

An important setting for the SU($N$) FH model is ultracold alkaline-earth-atoms (AEAs), where a highly precise $\rm{SU(}$$N)$ symmetry arises from the near-perfect decoupling of nuclear spins from the electronic structure due to their closed shells~\cite{Wu2003, Cazalilla2009, Gorshkov2010, Cazalilla2014, Stellmer2014}. Using fermionic isotopes of \ce{^{87}Sr} and \ce{^{173}Yb}, recent ultracold atom experiments have successfully observed Mott insulating states~\cite{Shintaro2012, Hofrichter2016, Tusi2022}, nearest-neighbor (NN) antiferromagnetic correlations~\cite{Ozawa2018, Taie2022}, as well as measured the equation of state in the $\rm{SU(6)}$ FH model~\cite{Pasqualetti2023}.

While the intricate magnetic structures of $\rm{SU(}$$N)$ Heisenberg magnets have been studied with increased interest~\cite{Manmana2011, Corboz2011, Nataf2014, Nataf2016, Romen2020}, the physics of doped $\rm{SU(}$$N)$ Mott insulators remains widely unexplored. Studying doped Mott insulators with higher spin symmetries promises novel insights into the competition between spin and motional degrees of freedom, possibly helping to unravel the microscopic nature of hole pairing in strongly correlated electronic systems. 

In this article, we utilize tensor network methods to study the singly doped $\rm{SU(3)}$ $t$-$J$ model, both in and out of equilibrium. We establish the appearence of magnetic polarons consisting of chargons and flavor defects (\rm{SU(3)} spinons), whose dynamics is constrained to a single effective dimension along the ordered diagonal. We demonstrate how this sub-dimensional phenomenology is qualitatively captured within non-linear geometric string theory after including both chargon and spinon fluctuations of the magnetic polaron. Our results can be directly probed with ultracold AEAs paired with quantum gas microscopes with single-site resolution~\cite{Miranda2015, Miranda2017, Okuno2020}, paving the way towards exploring a wealth of exotic physics in doped $\rm{SU(}$$N)$ symmetric systems.

This article is organized as follows. In Sec.~\ref{sec:model}, we introduce the doped SU(3) $t$-$J$ model and present ground state results for single hole dopant. In Sec.~\ref{sec:NLST}, we describe the magnetic polaron using semi-analytical methods (non-linear geometric string theory). In Sec.~\ref{sec:Dynamics}, we analyze quench dynamics of the magnetic polaron both using numerical and semi-analytical methods, before summarizing and discussing our results in Sec.~\ref{sec:Discussion}.

\section{Ground state}\label{sec:model} In the limit of strong on-site repulsion, the hole-doped 2D $\rm{SU(3)}$ symmetric FH model as realized by AEAs in sufficiently deep lattices at density $\braket{n}\le1$ reduces to the $\rm{SU(3)}$ symmetric $t$-$J$ model on the square lattice. Neglecting three-site terms~\footnote{At low doping as considered here, three-site terms are expected to have only a minor effect on the appearing physics~\cite{Ammon1995}. However, at finite doping, a systematic analysis of the terms may be necessary to draw direct connections to ultracold atom experiments.}, the corresponding Hamiltonian reads
\begin{equation}
    \begin{aligned}
    \hat{\mathcal{H}} =& -t \sum_{\braket{\mathbf{i,j}}, \alpha}  \hat{\mathcal{P}} \left( \hat{c}^{\dagger}_{\mathbf{i},\alpha} \hat{c}_{\mathbf{j},\alpha}^{\vphantom\dagger} + \text{h.c.} \right) \hat{\mathcal{P}} \\ &+ \frac{J}{2} \sum_{\braket{\mathbf{i,j}}} \left(\sum_{\alpha, \beta} \hat{\mathcal{P}} \hat{c}^{\dagger}_{\mathbf{i},\alpha} \hat{c}_{\mathbf{i},\beta}^{\vphantom\dagger} \hat{c}^{\dagger}_{\mathbf{j},\beta} \hat{c}_{ \mathbf{j},\alpha}^{\vphantom\dagger}\hat{\mathcal{P}} - \hat{n}_{\mathbf{i}} \hat{n}_{\mathbf{j}} \right).
    \end{aligned}
    \label{eq:H}
\end{equation}
Here, $\hat{c}^{\dagger}_{\mathbf{i},\alpha}$ ($\hat{c}_{\mathbf{i},\alpha}$) are creation (annihilation) operators of flavor $\alpha = \{\text{R,G,B}\}$ on site $\mathbf{i}$, $\braket{\mathbf{i}, \mathbf{j}}$ denotes nearest neighbors on the 2D square lattice, $\hat{n}_\mathbf{i}$ are the local particle densities, and $\hat{\mathcal{P}}$ is the Gutzwiller operator projecting out states with local occupancy $>1$. Note that for $N=2$, the second part of Eq.~\eqref{eq:H} reduces to $J \sum_{\braket{\mathbf{i}, \mathbf{j}}}\left( \hat{\mathbf{S}}_{\mathbf{i}}\cdot \hat{\mathbf{S}}_{\mathbf{j}} - \hat{n}_{\mathbf{i}}\hat{n}_{\mathbf{j}}/4\right)$ as in the usual $t$-$J$ model, with $\hat{\mathbf{S}}_{\mathbf{i}}$ the spin-$1/2$ operators~\cite{Auerbach1998}. For a single hole, the density-density interaction in Eq.~\eqref{eq:H} merely leads to a constant energy shift (up to boundary effects) and is neglected in the following, see also Appendix~\ref{sec:tJapp}.

We simulate the ground state of the one-hole doped $\rm{SU(3)}$ $t$-$J$ model, Eq.~\eqref{eq:H}, using the density matrix renormalization group (DMRG)~\cite{Schollwoeck_DMRG, SchollwoeckDMRG2, WhiteDMRG, hubig:_syten_toolk, hubig17:_symmet_protec_tensor_networ}. 
\begin{figure}
\centering
\includegraphics[width=0.9\columnwidth]{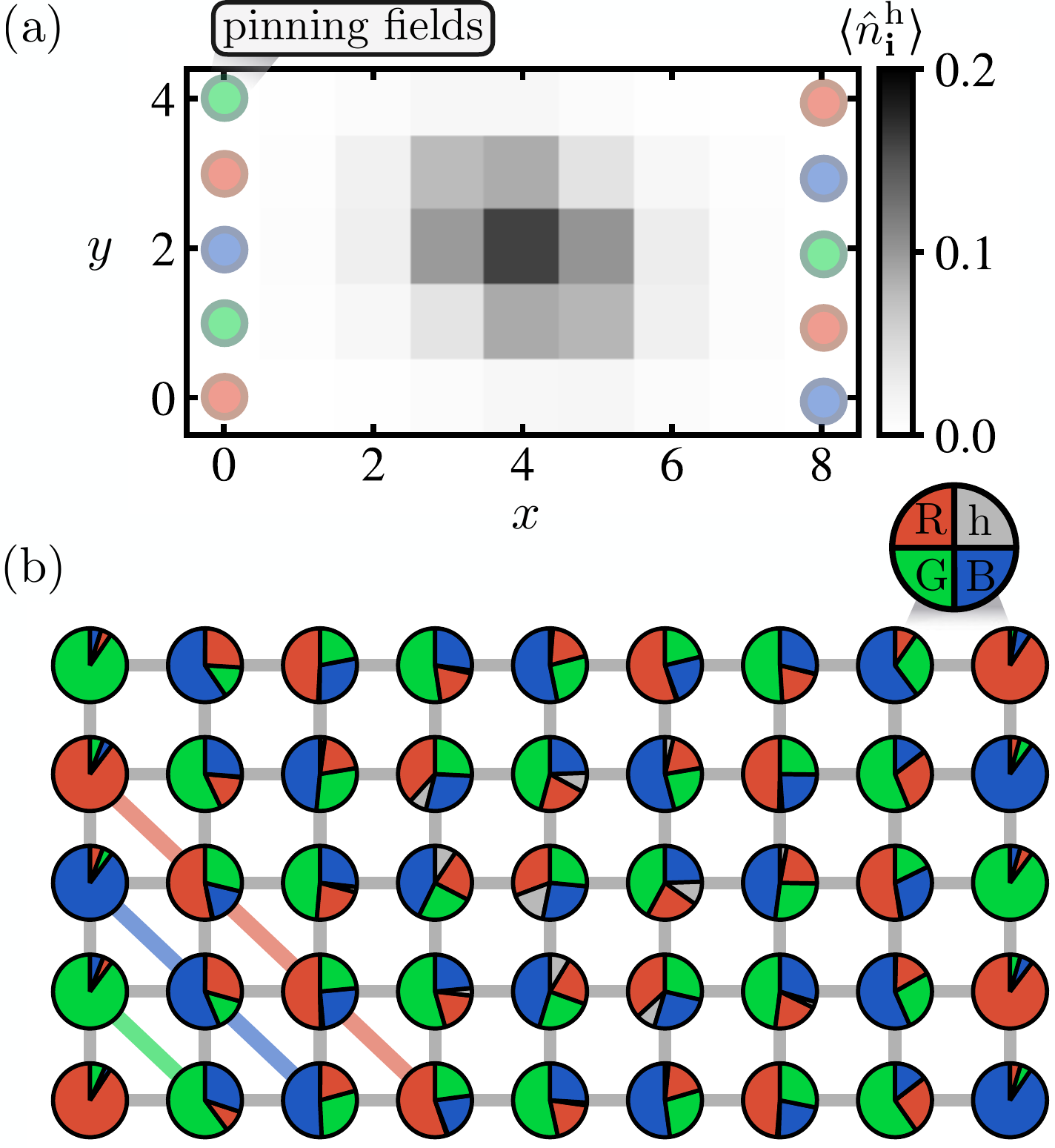}
\caption{\textbf{Single hole ground state.} (a) DMRG results of the hole density distribution $\braket{\hat{n}^h_{\mathbf{i}}}$ for a single hole (missing red atom) doped into a magnetic background pinned at the edges (indicated by the colored circles along the edges), for a $L_x \times L_y = 9 \times 5$ system with open boundaries, $t/J = 1.5$, and pinning fields $\mu_p/J = 1$. The hole distribution is anisotropic, such that the hole spreads further along the diagonal spin stripes than perpendicular to them. The full on-site moment distributions $\braket{\hat{n}_{\mathbf{i}}^f}$, $f= \text{h,R,G,B}$, are shown in gray, red, green and blue, respectively, in (b). The 3-SL diagonal stripe order pinned by the local chemical potentials on the boundaries is indicated by the solid colored lines in (b).}
\label{fig:density}
\end{figure}
We implement separate $\rm{U(1)}$ particle conservation symmetries for each flavor, and simulate systems of size $L_x \times L_y = 9\times 5$. Using bond dimensions of $\chi = 5000$, we carefully ensure convergence of all observables, see also Appendix~\ref{sec:conv_gs}. In the undoped case, open boundary conditions (OBC) have been identified as crucial to observe a diagonally striped ground state in numerically available systems sizes and were argued to capture the physics of the thermodynamic limit more accurately than periodic boundaries~\cite{Bauer2012}. 
Therefore, also in the case of the hole-doped $\rm{SU(3)}$ $t$-$J$ model, we focus on OBC along both $x-$ and $y-$ directions in the following. For a more detailed discussion, we refer to Appendix~\ref{sec:BCs}.

\begin{figure*}
\centering
\includegraphics[width=\textwidth]{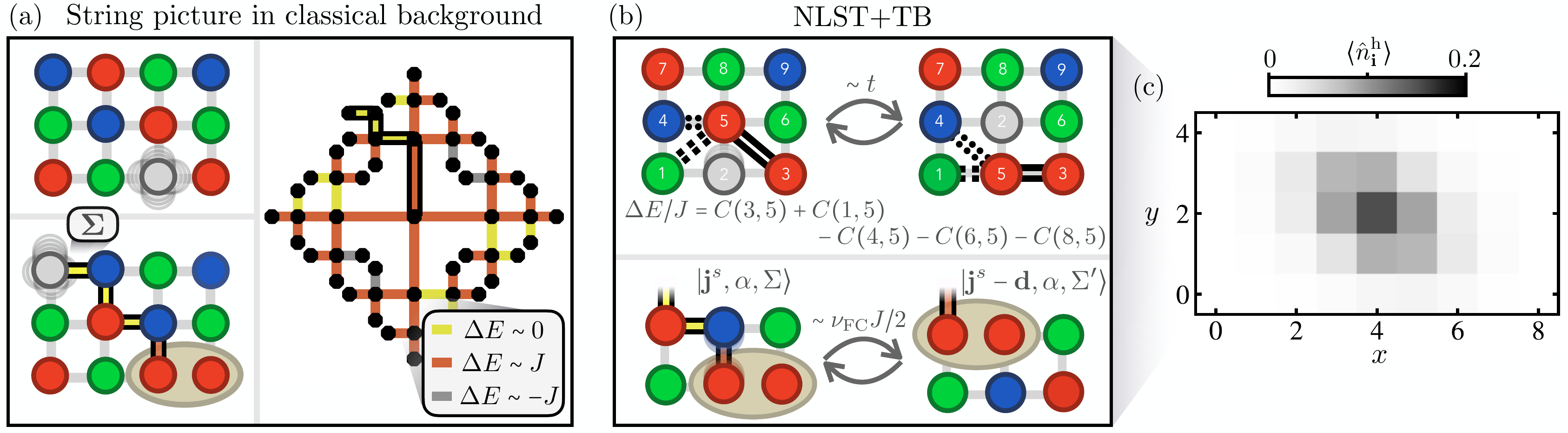}
\caption{\textbf{Non-linear string theory.} (a) Illustration of hole motion through a classically ordered $\rm{SU(3)}$ background. When hopping, the hole leaves behind a string $\Sigma$ of displaced spins. The first hop is confining due to positive spin-spin correlations along the diagonal -- leading to the formation of a spinon, see the ocher ellipse in the lower left panel. Subsequent hops along the diagonals, however, lead to classically degenerate configurations and paths of no additional energy cost (yellow). The full energy landscape is illustrated on the Bethe lattice in the right panel. (b) We describe chargon fluctuations within the FSA. String state energies are determined by a sum of particle exchange correlations $C(\mathbf{i}, \mathbf{j})$ of the undoped ground state, where $\mathbf{i},\mathbf{j}$ become neighbors after the hole is displaced through the string (upper panel). Spinon fluctuations are included by considering dominant off-diagonal couplings (lower panel), leading to an effective 1D spinon Hamiltonian on the diagonal. (c) The resulting hole density distribution when including both chargon and spinon fluctuations, matching DMRG results presented in Fig.~\ref{fig:density}.}
\label{fig:NLST}
\end{figure*}

By introducing local chemical potentials $-\mu_p \sum_{i\in \text{edge}} \hat{n}_{\alpha_{\mathbf{i}}, \mathbf{i}}$ at the short edges of the system, we explicitly break the $\rm{SU(3)}$ symmetry and pin a 3-sublattice (3-SL) stripe order along the diagonal, indicated by colored lines in Fig.~\ref{fig:density}~(a). We here fix $\mu_p/J = 1$; however, in Appendix~\ref{sec:pinningApp} we show that magnetic correlations are independent of the pinning potential in the bulk of the system. We introduce a single hole into the system by removing a particle corresponding to the flavor of the central site $\mathbf{i}_0$ in a (classically) ordered background. For the pinning shown in Fig.~\ref{fig:density}, this corresponds to the symmetry sector $N_{\text{G}} = N_{\text{B}} = L_xL_y/3$, $N_{\text{R}} = L_x L_y/3 - 1$, where $N_{\alpha}$ is the total particle number of flavor $\alpha =$ R (red), G (green), B (blue). 

The hole density distribution of the ground state determined by DMRG, $\braket{\hat{n}^h_{\mathbf{i}}} = 1 - \sum_{\alpha} \braket{\hat{n}^{\alpha}_ \mathbf{i}}$, is shown in Fig.~\ref{fig:density}~(a) for $t/J = 1.5$. The hole density features a pronounced anisotropy, whereby its distribution has enhanced weight on the diagonals aligning with the pinned order. On the other hand, the hole density is suppressed on the anti-diagonals, i.e, directions that are perpendicular to the 3-SL order. This is corroborated in Fig.~\ref{fig:density}~(b), where the full on-site moments $\braket{\hat{n}^f_{\mathbf{i}}}$, for $f = $ h (hole), R (red), G (green) and B (blue) are shown. Importantly, the delocalization of the hole only slightly disturbs the magnetic background in its vicinity, leaving the overall 3-SL order intact, as illustrated by the solid colorful lines in the lower left corner of Fig.~\ref{fig:density}~(b).

\section{Geometric string theory}\label{sec:NLST} In the following, we describe the doped hole in the 3-SL background using non-linear geometric string theory (NLST)~\cite{GrusdtX, Grusdt_strings} and establish the formation of a sub-dimensional magnetic polaron whose motion is predominantly aligned with the ordered background. The starting point is a parton representation of the $\rm{SU(3)}$ $t$-$J$ model, where the creation and annihilation operators are decomposed into bosonic chargons ($\hat{h}_{\mathbf{i}}$) and fermionic spinons ($\hat{f}_{\mathbf{i},\alpha}$), $\hat{c}_{\textbf{i},\alpha} = \hat{h}_{\mathbf{i}}^{\dagger} \hat{f}_{\mathbf{i},\alpha}^{\vphantom\dagger}$. The spinon label $\alpha$ corresponds to the flavor of the particle that has been removed. The single occupancy constraint in the $t$-$J$ model is ensured via $\sum_{\alpha} \hat{f}^{\dagger}_{\alpha, \mathbf{i}} \hat{f}_{\alpha, \mathbf{i}}^{\vphantom\dagger} + \hat{h}^{\dagger}_{\mathbf{i}} \hat{h}_{\mathbf{i}}^{\vphantom\dagger} = 1$ for all $\mathbf{i}$. 

We describe the magnetic polaron within the geometric string basis: By doping a single hole at position $\mathbf{j}^s$ into the ground state $\ket{\Psi_0}$ of the undoped $\rm{SU(3)}$ Heisenberg Hamiltonian, we define the state $\ket{\mathbf{j}^s, \alpha, \Sigma=0} = \hat{c}_{\alpha, \mathbf{j}^s} \ket{\Psi_0}$. To describe the partons, we work in the regime $t/J \gg 1$, where fluctuations of the chargon and the ordered background approximately decouple. In a first step, we fix the initial hole position $\mathbf{j}^s$, and describe the fast chargon fluctuations on time scales $\sim 1/t$. Motivated by the separation of energy scales, we work in the frozen spin approximation (FSA): When the hole fluctuates, the background spins are displaced and their positions change, however their quantum state is assumed to remain unaffected. This generalizes the notion of squeezed space~\cite{OgataShiba, Zaanen2001, Kruis2004} to two dimensions. By displacing particles in real space, the hole motion changes the underlying geometry of the lattice in squeezed space, whereby nearest neighbor (NN) pairs in squeezed space can become next-nearest neighbor (NNN) or even larger-distance pairs. String states are defined by $\ket{\mathbf{j}^s, \alpha, \Sigma} = \hat{G}_{\Sigma} \ket{\mathbf{j}^s, \alpha, \Sigma=0}$, where the string operator $\hat{G}_{\Sigma} = \prod_{\braket{\mathbf{i}, \mathbf{j}}\in \Sigma} \left( \hat{h}^{\dagger}_{\mathbf{i}} \hat{h}_{\mathbf{j}}^{\vphantom\dagger} \sum_{\alpha} \hat{f}^{\dagger}_{\mathbf{j},\alpha} \hat{f}_{\mathbf{i}, \alpha}^{\vphantom\dagger} \right)$ displaces the background spins along string $\Sigma$.

The string states and the tunnel coupling between them can be mapped to a Bethe lattice, illustrated in Fig.~\ref{fig:NLST}~(a) for a classical spin background. In particular, each string of displaced particles $\Sigma$ can be associated with a corresponding potential energy on the Bethe lattice, $E(\mathbf{j}^s, \Sigma)$~\footnote{Note that in the $\rm{SU(2)}$ case, where the background is ordered according to a 2-SL structure, all directions are equally confining, i.e., the string energy only depends on the absolute length of the string, $E(\Sigma) = E(|\Sigma|)$.}. Alignment of flavors along the diagonals surrounding the hole leads to an energy penalty $\Delta E \sim J$ when the hole leaves its initial position and moves by one lattice site, leaving behind a flavor defect (spinon) at $\mathbf{j}^s$ [see the ocher ellipse in the lower left panel of Fig.~\ref{fig:NLST}~(a)]. After its initial hop, however, the energy landscape for subsequent chargon motion loses its isotropy. In particular, paths along the diagonally ordered background merely lead to local flavor exchanges along the string, which is classically degenerate with the initial 3-SL order. This, in turn, leads to the existence of string segments of no additional energy cost, illustrated by yellow paths on the Bethe lattice in the right panel of Fig.~\ref{fig:NLST}~(a). Directions perpendicular to the diagonal stripe order, in contrast, are associated with linearly growing magnetic energy (linear confinement), shown by red paths in Fig.~\ref{fig:NLST}~(a). Negative energy differences (gray lines) result from loop effects.

\begin{figure*}
\centering
\includegraphics[width=\textwidth]{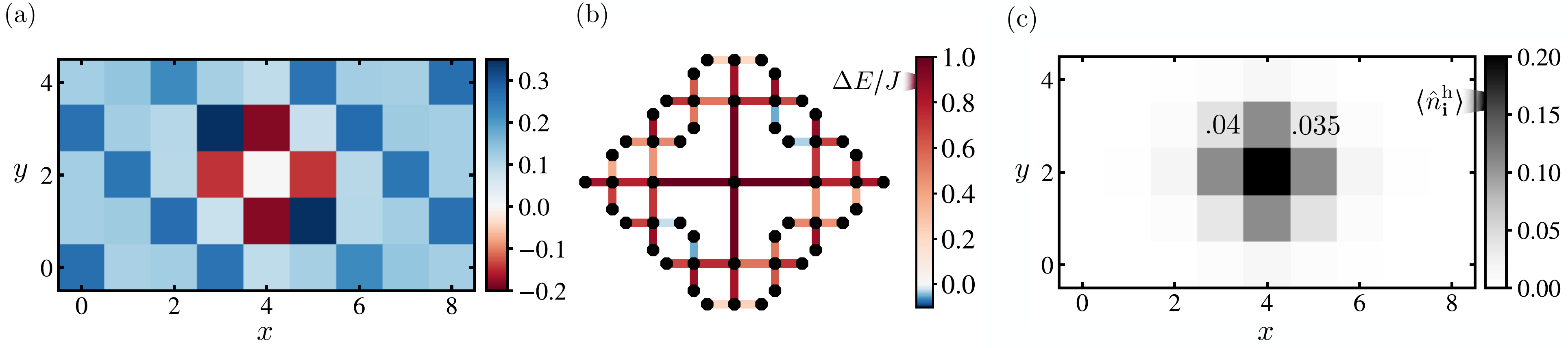}
\caption{\textbf{Chargon fluctuations within NLST.} (a) Correlations $C(\mathbf{i},\mathbf{j})$, Eq.~\eqref{eq:corr}, for a fixed reference site $\mathbf{i} = [4,2]$ calculated with DMRG with the same pinning fields as described in the main text. These correlations are the essential ingredient for calculating the energy of string states on the Bethe lattice, presented for lattice depth $d=3$ in (b). The structure follows the classical picture illustrated in Fig.~\ref{fig:NLST}~(a) in the main text. Finite size effects are taken into account by cutting off sites on the Bethe lattice that do not lie within the finite system's frame. (c) Density distribution in real space after diagonalizing the hopping Hamiltonian on the Bethe lattice, Eq.~\eqref{eq:Hjf}. Though the anisotropic energy distribution on the Bethe lattice leads to slightly larger hole densities on the diagonals compared to the anti-diagonals, differences are small (see the numerical values).}
\label{fig:nlst}
\end{figure*}

In the case of a classical background, string states $\ket{\mathbf{j}^s, \alpha, \Sigma}$ are mutually orthonormal except for special loop configurations which restore the ordered background, known as Trugman loops~\cite{Trugman1988, Grusdt_strings, GrusdtX}. In the case of a (classical) diagonally striped background, these configurations involve at least 12 string segments (corresponding to three loops around a square plaquette), and are hence negligible compared to the exponential number of string states. Due to strong diagonal-order correlations in the undoped $\rm{SU(3)}$ $t$-$J$ model (see Sec.~\ref{sec:chargon}), we expect that the approximation of mutual orthonormality of string states remains accurate away from the classical limit.

We include quantum fluctuations in two stages: First, we describe fluctuations of the chargon moving in a frozen spin background. In a second step, we include fluctuations of the spinon through a tight-binding description. The formalism is summarized in Fig.~\ref{fig:NLST}~(b) and explained in detail in the following.

\subsection{Chargon motion}
\label{sec:chargon}
Our first step to go beyond the classically ordered magnetic background is to use the FSA: Upon creation of the hole, the background spins are labeled according to their original positions. When the hole moves, the particles are displaced and their positions change, resulting in energies $E(\mathbf{j}^s, \Sigma)$ of string states $\ket{\mathbf{j}^s, \alpha, \Sigma}$. The crucial ingredient in order to describe fluctuations of the chargon around its initial position $\mathbf{j}^s$ within NLST are magnetic correlations of the undoped system,
\begin{equation}
    C(\mathbf{i},\mathbf{j}) = \sum_{\alpha, \beta} \braket{\Psi_0|\hat{\mathcal{P}} \hat{c}^{\dagger}_{\alpha, \mathbf{i}} \hat{c}_{\beta, \mathbf{i}} \hat{c}^{\dagger}_{\beta, \mathbf{j}} \hat{c}_{\alpha, \mathbf{j}}\hat{\mathcal{P}}|\Psi_{0}}/2,
    \label{eq:corr}
\end{equation}
where $\ket{\Psi_0}$ is the ground state with one particle per site. Fig.~\ref{fig:nlst} shows correlations for a fixed reference site $\mathbf{i} = [x=4,y=2]$ in the center of the finite-size system we study with DMRG. 
Following the 3-SL order of the ground state, sites are correlated positively along every third diagonal; nearest neighbor correlations, in contrast, show strong negative signals. When a hole is added at initial position $\mathbf{j}^s$ and then hops away, assuming a frozen spin background it reshuffles the particles in its vicinity, leading to an energy cost that can directly be evaluated from correlations given by Eq.~\eqref{eq:corr}. 

\begin{figure*}
\centering
\includegraphics[width=\textwidth]{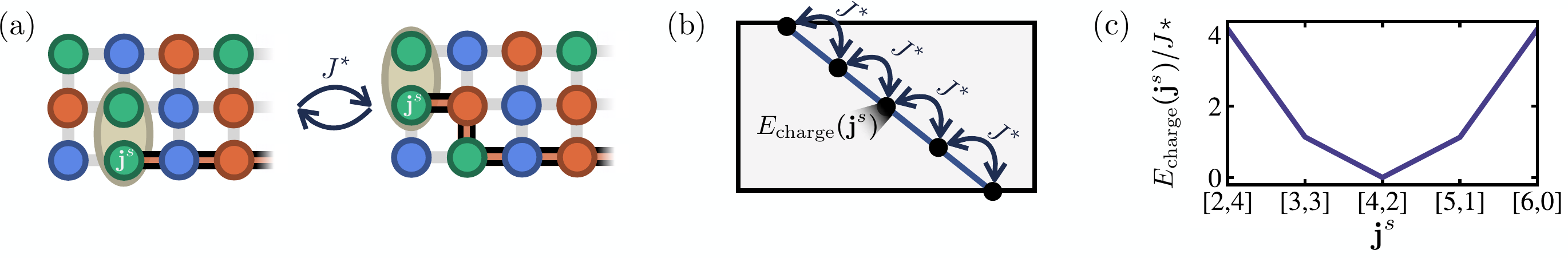}
\caption{\textbf{Spinon tight-binding description.} Exchange processes as appearing in the Hamiltonian of the $\rm{SU(3)}$ $t$-$J$ model lead to hopping processes of the spinon, as illustrated in (a). Dominant contributions come from diagonal spinon hopping events, (b), resulting in a string state $\ket{\Sigma', \mathbf{j}^s + \mathbf{d}}$ of length $|\Sigma'| = |\Sigma| \pm 2$. The on-site energy $E_{\text{charge}}(\textbf{j}^s)$ defines an effective potential seen by the spinon, (c), here shown for $\nu_{\text{FC}} = 0.1$.}
\label{fig:ftb}
\end{figure*}

For instance, if a hole initially placed in the center $\mathbf{j}^s = [4,2]$ moves up by one lattice site, the corresponding change in magnetic energy is given by $\Delta E/J = C(\mathbf{i} = [3,2], \mathbf{j} = [4,3]) + C(\mathbf{i} = [5,2], \mathbf{j} = [4,3]) + C(\mathbf{i} = [4,1], \mathbf{j} = [4,3]) - C(\mathbf{i} = [4,4], \mathbf{j} = [4,3]) - C(\mathbf{i} = [5,3], \mathbf{j} = [4,3]) - C(\mathbf{i} = [3,3], \mathbf{j} = [4,3])= 0.97$. Using correlations Eq.~\eqref{eq:corr}, all energies on the Bethe lattice are calculated this way, depicted in Fig.~\ref{fig:nlst}~(b) for $\mathbf{j}^s = [4,2]$. Note that this includes the effect of open boundary conditions within NLST. Additionally, if the hole moves outside the finite system's frame, the site is cut off from the Bethe lattice [in Fig.~\ref{fig:nlst}~(b), a Bethe lattice depth of $d=3$ is considered. The two sites corresponding to strings (up, up, up) and (down, down, down) lie outside the frame and are thus cut off]. Energies of string configurations (diagonal matrix elements of a string configuration with itself) using the system's correlations follow the structure of a classical background as illustrated in Fig.~\ref{fig:NLST}~(a) in the main text, whereby string energies along the diagonal are lower compared to other paths. 

The Hamiltonian of the chargon for an initial hole position $\mathbf{j}^s$ is then expressed within the geometric string basis on the Bethe lattice, 
\begin{equation}
\begin{aligned}
    \hat{\mathcal{H}}_{\text{charge}}(\mathbf{j}^s, \alpha) =-t &\sum_{\braket{\Sigma, \Sigma'}} \ket{\mathbf{j}^s, \alpha, \Sigma}\bra{\mathbf{j}^s, \alpha, \Sigma'} + \text{ h.c.} \\ +& \sum_{\Sigma} E(\mathbf{j}^s, \Sigma) \ket{\mathbf{j}^s, \alpha, \Sigma}\bra{\mathbf{j}^s, \alpha, \Sigma}.
\end{aligned}
\label{eq:Hjf}
\end{equation}
The Hamiltonian Eq.~\eqref{eq:Hjf} is diagonalized, yielding
\begin{equation}
    \ket{\Psi_{\text{charge}}(\mathbf{j}^s, \alpha)} = \sum_{\Sigma} \psi^{\text{charge}}_{\mathbf{j}^s, \Sigma} \ket{\mathbf{j}^s, \alpha, \Sigma}
    \label{eq:nlst_gs}
\end{equation}
as its ground state with eigenenergy $E_{\text{charge}}^{0}(\mathbf{j}^s)$. Mapping the hole density back to real space, $\braket{\hat{n}^h_{\mathbf{j}}}= \sum_{\Sigma \in Q_{\mathbf{j}}} |\psi^{\text{charge}}_{\mathbf{j}^s, \Sigma}|^2$ where $Q_{\mathbf{j}}$ is the set of paths leading from a hole initially at $\mathbf{j}^s$ to be located at $\mathbf{j}$, results in the density distribution shown in Fig.~\ref{fig:nlst}~(c) for $\mathbf{j}^s = [4,2]$. Though the anisotropic energy distribution on the Bethe lattice leads to slightly larger hole densities on the diagonals compared to the anti-diagonals, differences are small [see the numerical values in Fig.~\ref{fig:nlst}~(c)]. 

Comparing the result to the densities as acquired from DMRG calculations, this suggests that the magnetic polaron itself (built up from fast chargon fluctuations centered around the fixed spinon) is only slightly influenced by the anisotropic energies on the Bethe lattice, but instead the motion of the composite chargon-spinon object induces the observed anisotropy. 

\subsection{Spinon fluctuations}
Instead, in the following we demonstrate that is the effect of spinon fluctuations on time scales $\propto 1/J$ that lead to the observed alignment along the diagonal in Fig.~\ref{fig:density}~(c), which we include on top of chargon fluctuations by using a tight-binding description of spinon motion (NLST+TB). Concretely, we consider off-diagonal couplings $J_s(\mathbf{j}^s, \mathbf{j}^{s'}; \Sigma, \Sigma') = \braket{\mathbf{j}^{s'}, \alpha,\Sigma'|\hat{\mathcal{H}}|\mathbf{j}^{s},\alpha, \Sigma}$ within the geometric string basis construction to describe spinon fluctuations. In the case of the $\rm{SU(2)}$ symmetric $t$-$J$ model, major contributions are given by next-nearest neighbor (NNN) spinon hopping processes isotropically in all spatial directions. In contrast, in the $\rm{SU(3)}$ $t$-$J$ model, dominant contributions are restricted to diagonal NNN spinon hopping processes along the 3-SL order. This is owing to the Hamiltonian's $\rm{U(1)}^{\otimes N}$ particle conservation symmetry: applying $\hat{\mathcal{H}}$ to string states $\ket{\mathbf{j}^{s},\alpha, \Sigma}$ conserves the flavor of the removed particle $\alpha$.

Fig.~\ref{fig:ftb}~(a) illustrates the process. For a given string configuration $\ket{\mathbf{j}^s, \alpha = \text{R}, \Sigma}$, exchange of two neighboring particles (here given by the green and red flavors at the left edge of the central leg) leads to a string configuration $\ket{\mathbf{j}^s - \mathbf{e}_x + \mathbf{e}_y, \alpha = \text{R}, \Sigma'}$, with a resulting string length $|\Sigma'| = |\Sigma| + 2$ and unit vectors $\mathbf{e}_x, \mathbf{e}_y$. More generally, the Hamiltonian couples off-diagonal string states $\braket{\mathbf{j}^s, \alpha, \Sigma|\hat{\mathcal{H}}|\mathbf{j}^s \pm \mathbf{d}, \alpha, \Sigma'}$ with $|\Sigma'| = |\Sigma| \pm 2$ and $\mathbf{d} = \mathbf{e}_x - \mathbf{e}_y$ pointing along the diagonal stripe order, see also Fig.~\ref{fig:NLST}~(b).

\begin{figure*}
\centering
\includegraphics[width=0.73\textwidth]{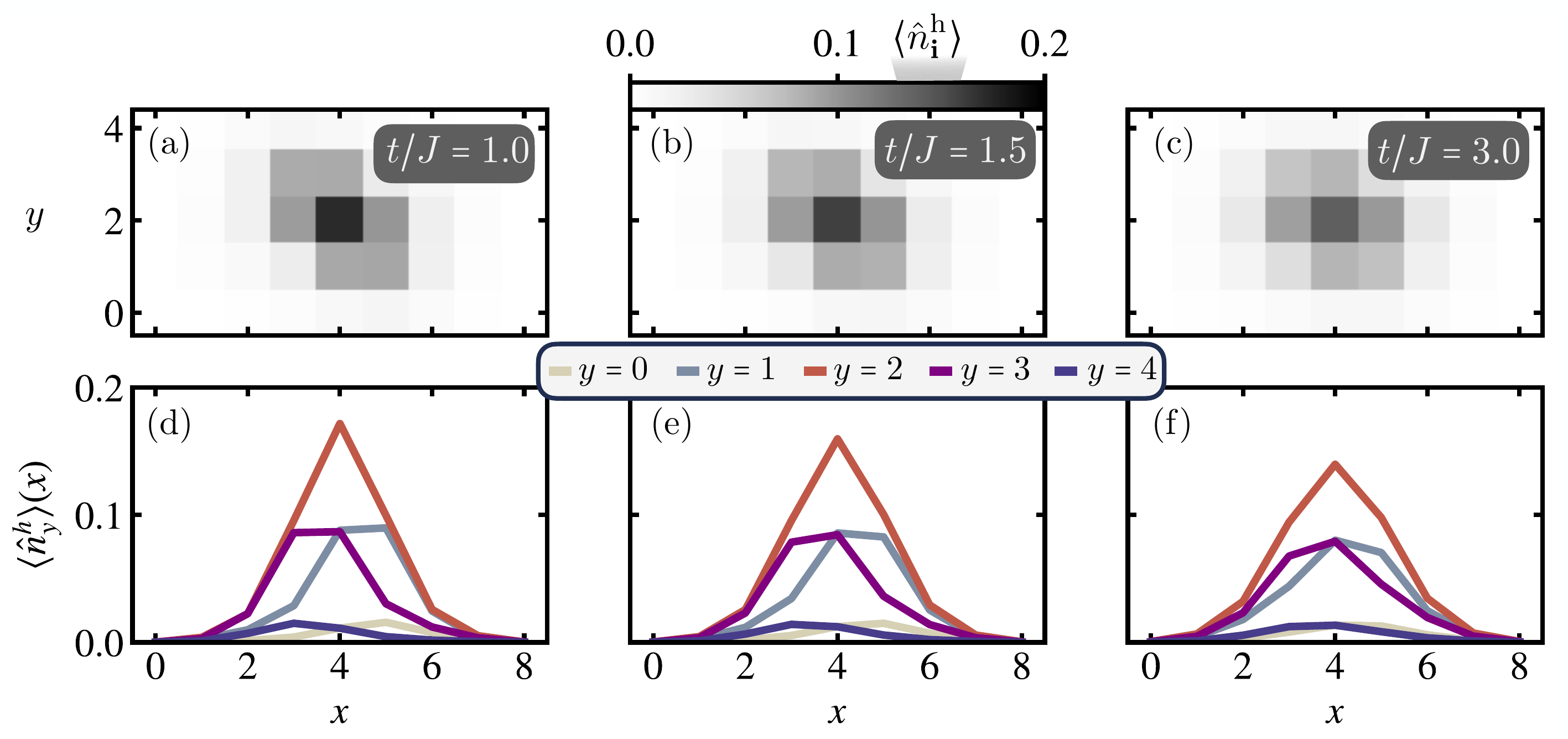}
\caption{\textbf{Varying $\mathbf{t/J}$.} (a)-(c) 2D hole density distributions for $t/J = 1.0, 1.5, 3.0$, respectively, calculated with DMRG for $L_x \times L_y = 9\times 5$ systems with OBC. (d)-(f) Hole densities along the $x$-direction, for each of the five legs. When increasing $t/J$ (i.e. increasing the ratio of hole fluctuations compared to magnetic coupling strength), the anisotropy in the density is observed to monotonously decrease.}
\label{fig:vart}
\end{figure*}

Owing to the finite overlap of string states with different initial hole positions $\mathbf{j}^s$, the effective spinon hopping is given by
\begin{equation}
    J^*(\mathbf{j}^s, \mathbf{j}^{s'}) = \sum_{\Sigma, \Sigma'} J_s(\mathbf{j}^s, \mathbf{j}^{s'}; \Sigma, \Sigma') \psi^{\text{charge} *}_{\mathbf{j}^{s'}, \Sigma'} \psi^{\text{charge}}_{\mathbf{j}^s, \Sigma}.
    \label{eq:nuFC}
\end{equation}
In the classical limit and for large system sizes, $J_s(\mathbf{j}^s, \mathbf{j}^{s'}, \Sigma, \Sigma') = J/2$ if a single particle exchange relates the two string states $\ket{\mathbf{j}^{s}, \alpha, \Sigma}$ and $\ket{\mathbf{j}^{s'}, \alpha,\Sigma'}$. As the exact evaluation of $J_s(\mathbf{j}^s, \mathbf{j}^{s'}, \Sigma, \Sigma')$ is cumbersome, we approximate $J^*(\mathbf{j}^s, \mathbf{j}^{s'}) \approx \nu_{\text{FC}} J/2$, where we treat the Franck-Condon factor $\nu_{\text{FC}}$ as an effective fit parameter of the geometric string theory. In the limit of weak coupling, $t\ll J$, the Franck-Condon factor approaches $\nu_{\text{FC}} = 0$. In the strong coupling regime, $t \gg J$, $\nu_{\text{FC}} \rightarrow 0.5$~\cite{Grusdt_strings}.

We model diagonal hopping of the heavy polaron by an effectively 1D tight-binding system, with hopping parameter $J^* = \nu_{\text{FC}} J/2$ and on-site energies $E_{\text{charge}}^{0}(\mathbf{j}^s)$ calculated via NLST, cf. Eq.~\eqref{eq:nlst_gs}, with $\mathbf{j}^s$ lying on the diagonal $\mathcal{D}$ that includes the central site -- illustrated in Fig.~\ref{fig:ftb}~(b) and (c). The Hamiltonian is given by
\begin{equation}
\begin{aligned}
    \hat{\mathcal{H}}_{\text{spinon}} = J^* \Big( \sum_{\braket{\mathbf{i},\mathbf{j}}\in \mathcal{D}} c_{\textbf{i}}^{\dagger} c_{\textbf{j}} &+ \text{h.c.} \Big) \\ &+ \sum_{\mathbf{j}\in \mathcal{D}} E^{0}_{\text{charge}}(\mathbf{j}) c_{\textbf{j}}^{\dagger} c_{\textbf{j}},
\end{aligned}
\end{equation}
yielding coefficients $\psi^{\text{spinon}}_{\mathbf{j}^s}$ for spinon positions $\mathbf{j}^s$. We note that, as both $\nu_{\text{FC}}$ and $J$ are positive, the effective spinon hopping is positive, $J^*>0$, yielding a dispersion minimum of the spinon at $k=\pi$ reminiscent to 2D quantum magnets~\cite{Grusdt_strings, Grusdt_tJ}. Finally, we combine charge and spinon parts by a plane-wave ansatz, arriving at
\begin{equation}
\begin{aligned}
    \ket{\Psi} = \sum_{\mathbf{j}^s \in \mathcal{D}} &\psi^{\text{spinon}}_{\mathbf{j}^s} \ket{\Psi_{\text{charge}}(\mathbf{j}^s, \alpha)}  \\&= \sum_{\mathbf{j}^s \in \mathcal{D}} \psi^{\text{spinon}}_{\mathbf{j}^s} \sum_{\Sigma} \psi^{\text{charge}}_{\mathbf{j}^s, \Sigma} \ket{\mathbf{j}^s, \alpha, \Sigma}.
    \label{eq:psi_gst_sm}
\end{aligned}
\end{equation}
The total hole distribution is given by a weighed sum with coefficients $|\psi^{\text{spinon}}_{\mathbf{j}^s}|^2$ of hole distributions $|\psi^{\text{charge}}_{\mathbf{j}^s,\Sigma}|^2$ for each mean chargon position $\mathbf{j}^s$, i.e., it is determined by mapping $|\psi^{\text{spinon}}_{\mathbf{j}^s} \psi^{\text{charge}}_{\mathbf{j}^s, \Sigma}|^2$ back to the original real space lattice. We note again that we treat $\nu_{\text{FC}}$ as an effective free parameter of the theory, matching DMRG results strikingly well for $\nu_{\text{FC}} = 0.1$, see Fig.~\ref{fig:NLST}~(c). This agreement corroborates the validity of NLST+TB, and supports the existence of a sub-dimensional magnetic polaron in the singly doped $\rm{SU(3)}$ $t$-$J$ model. Additional DMRG simulations presented in Fig.~\ref{fig:vart} further support this picture, whereby the anisotropy in the hole density is seen to increase for rising exchange interactions $J/t$, while a dominating hopping $J/t\ll 1$ leads to the formation of a broad, isotropic polaron cloud. Lastly, we note that while the Hilbert space spanned by the string states grows exponentially, a systematic cutoff of chargon states far away from their initial position on the Bethe lattice allows for an efficient calculation of hole density distributions. This ultimately allows us to make concrete comparisons between the phenomenological geometric string theory and finite-size numerical calculations. While the latter is only possible for small system sizes, our effective description provides strong evidence that the qualitative influence of the magnetic structure on the polaron distribution survives in the thermodynamic limit. This is further corroborated by large-scale calculations of the SU(3) FH model at one particle per site, which establish that the magnetic order is present also in large systems~\cite{Chunhan2023}.  

\section{Dynamics} \label{sec:Dynamics} To further study the behavior of the magnetic polaron in the $\rm{SU(3)}$ $t$-$J$ model, we analyze quenched hole dynamics, which is particularly accessible to ultracold atom experiments and has been probed for a single hole doped into an $\rm{SU(2)}$ AFM background~\cite{Ji2021}. Specifically, we analyze the hole's dynamics after doping it in the center $\mathbf{i}_0$ of the undoped ground state, i.e., at time $T=0$, the initial state is $\ket{\Psi(T=0)} = \ket{\mathbf{i}_0, \alpha, \Sigma = 0}$. We use the global subspace expansion~\cite{Yang2020} for a single time step, before switching to time dependent variational principle calculations~\cite{Paeckel_time}. During the time evolution, we track the Manhattan distance $\braket{\hat{x}_m}$ from $\mathbf{i}_0$ along the diagonal and anti-diagonal.

\begin{figure}
\centering
\includegraphics[width=0.94\columnwidth]{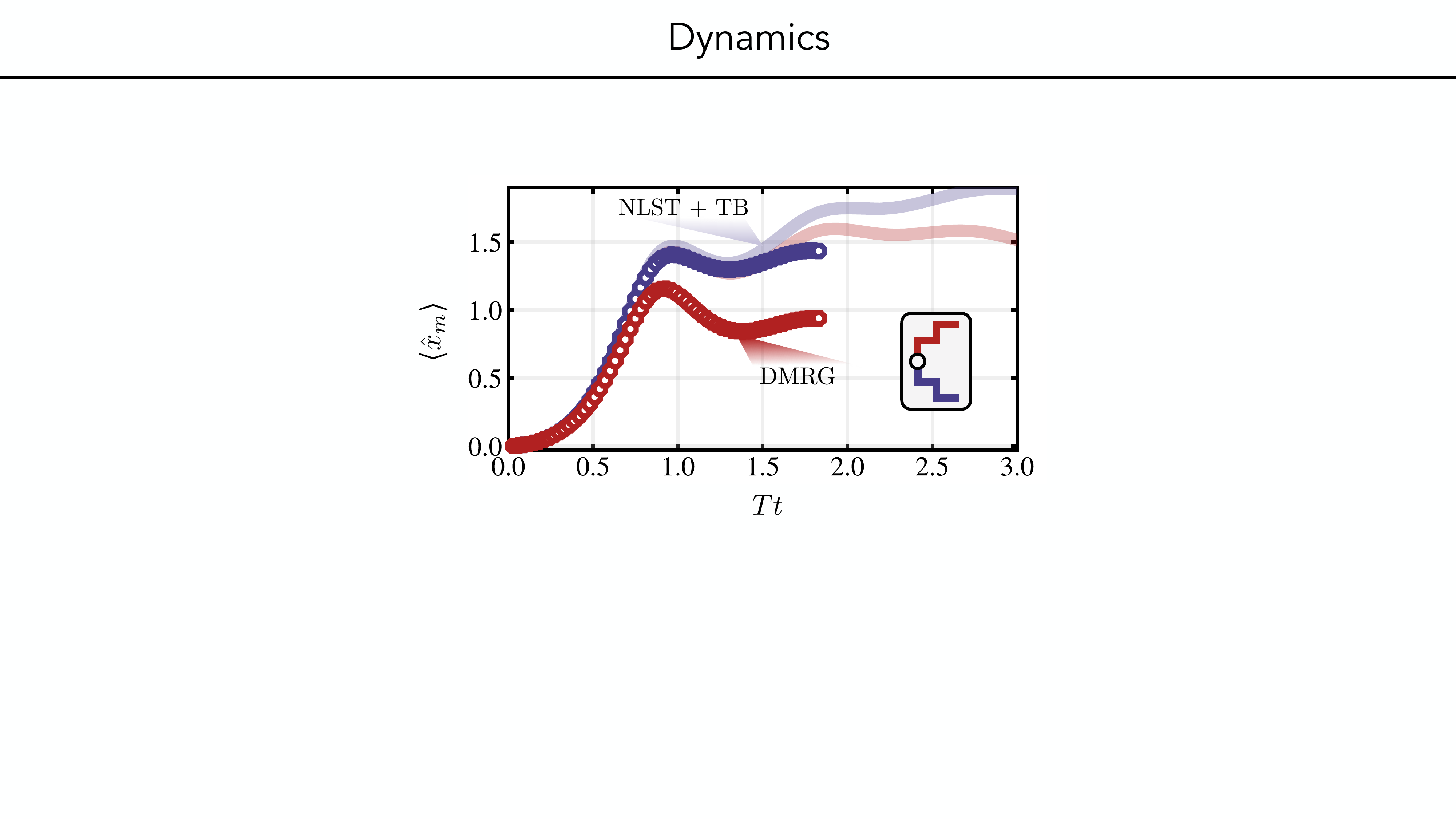}
\caption{\textbf{Hole dynamics.} Time evolution of an initially localized hole at $\mathbf{i}_0=[4,2]$. Mean Manhattan distances along the diagonal (anti-diagonal) are shown in blue (red) for a $L_x \times L_y = 9\times 5$ system. Light red and blue lines show NLST+TB results for an infinite system with $\nu_{\text{FC}} = 0.5$.}
\label{fig:dynamics}
\end{figure}

At short times, fast chargon fluctuations lead to a symmetric, ballistic expansion of the hole, see Fig.~\ref{fig:dynamics}. At times $Tt\sim 1$, a rapid slow down and apparent saturation of the hole's spread is observed, reminiscent of dynamics in the $\rm{SU(2)}$ $t$-$J$ model~\cite{Bohrdt2020_polaron, Hubig2020, Ji2021, Nielsen2022}. Here, it has been established that the hole's long-time dynamics is governed by spinon dynamics, i.e., by the motion of the heavy composite magnetic polaron itself. The strong splitting between diagonal and anti-diagonal distances appearing in the $\rm{SU(3)}$ system at times $Tt \sim 1$, shown in Fig.~\ref{fig:dynamics}, further underlines the role of spinon delocalization in the observed anisotropy.

We describe the dynamics of the magnetic polaron within geometric string theory by again assuming a product state ansatz with effective Hamiltonian 
\begin{equation}
    \hat{\mathcal{H}}_{\text{eff}} = \hat{\mathcal{H}}_{\text{charge}} + \hat{\mathcal{H}}_{\text{spinon}},
\end{equation}
such that dynamical properties are evaluated by calculating $\exp\left(-i\hat{\mathcal{H}}_{\text{eff}} T\right) \ket{\mathbf{j}^s, \alpha, \Sigma = 0}$; here, $\hat{\mathcal{H}}_{\text{spinon}}$ only acts on the spinon degree of freedom $\ket{\mathbf{j}^s}$, whereas $\hat{\mathcal{H}}_{\text{charge}}$ generates chargon dynamics for a given spinon position. As the MPS calculations are limited to small system sizes with OBC, we here focus on infinite systems (i.e. we do not consider the open boundaries when constructing the string basis) to make predictions of the dynamical formation and motion of the polaron in the thermodynamic limit. 

Results are shown in Fig.~\ref{fig:dynamics} with blue and red solid lines, where all qualitative features are in line with finite-size MPS simulations. At short times we find quantitative agreement. However, the anisotropy developing later in time is seen to be underestimated within the string theory, which is likely caused by significant finite size effects in the MPS dynamics (being particularly prominent as we are focusing on observables along the diagonal). In fact, motivated by the picture of a polaron effectively constrained to 1D, in the thermodynamic limit we expect a linear expansion (saturation) of $\hat{x}_m$ along the diagonal (anti-diagonal) at large times, which is, however, out of reach to simulate with current methods.  

\section{Discussion}\label{sec:Discussion} We have studied the one-hole doped $\rm{SU(3)}$ $t$-$J$ model both in- and out of equilibrium. In the ground state, we observed anisotropic hole delocalization, and established the formation of a sub-dimensional polaron by combining chargon and spinon fluctuations in an effective theory. This picture was further corroborated in calculations of the dynamics initiated from a localized hole, which can provide a direct probe of the polaron physics in $\rm{SU(}$$N)$ ultracold atom experiments once single-site resolution becomes available. In our setting of a doped $\rm{SU(3)}$ AFM, we have demonstrated how sub-dimensional excitations naturally emerge, reminiscent of mobility restricted fractons as appearing e.g. in three dimensional X-cube models~\cite{Vijay2016, Ma2017}.

Based on our study of a single hole, we propose that $\rm{SU(3)}$ AFMs on the square lattice at finite doping are described by weakly coupled Tomonaga-Luttinger (TL) liquids of bound spinon-chargon polarons along the diagonals, 
\begin{equation}
\begin{aligned}
    \hat{\mathcal{H}}_{\text{eff}} = \sum_{\mathcal{D}_i} &\hat{\mathcal{H}}_{\mathcal{D}_i} + \sum_{\mathcal{D}_i\neq \mathcal{D}_j} \hat{\mathcal{H}}_{\mathcal{D}_i,\mathcal{D}_j}^{\text{int}} \\ &+ \sum_{k_n} \sum_{\mathcal{D}_i\neq \mathcal{D}_j} t_{\mathcal{D}_i, \mathcal{D}_j} \hat{c}^{\dagger}_{k_n, \mathcal{D}_i} \hat{c}^{\vphantom\dagger}_{k_n, \mathcal{D}_j} + \text{h.c.},
    \end{aligned}
    \label{eq:LL}
\end{equation}
where $\hat{\mathcal{H}}_{\mathcal{D}_i}$ is the Hamiltonian of a 1D TL liquid~\cite{Voit1995} on the $i$'th diagonal $\mathcal{D}_i$, $\hat{\mathcal{H}}_{\mathcal{D}_i,\mathcal{D}_j}^{\text{int}}$ is the (weak) interaction between chains $\mathcal{D}_i$ and $\mathcal{D}_j$, and the last term describes particles with quasi-momentum $k_n$ hopping between the diagonals. In particular, in the absence of inter-chain couplings in Eq.~\eqref{eq:LL} we predict the appearance of power-law correlations of charges along the diagonal order, $\braket{\hat{n}^h_{x+\ell, y-\ell} \hat{n}^h_{x, y}} \propto \ell^{-\alpha}$, while correlations perpendicular to the stripe order are short-range with exponential decay, $\braket{\hat{n}^h_{x+\ell, y+\ell} \hat{n}^h_{x, y}} \propto e^{-\ell}$. Though inter-chain interactions are a relevant perturbation of the TL liquid, we still expect these scalings over intermediate length scales for finite couplings. \\

\textbf{Acknowledgments.---} We thank Tizian Blatz, Chunhan Feng, Simon F\"olling, Eduardo Ibarra-Garc\'ia-Padilla, Carlos Sa de Melo, Sebastian Paeckel, Richard Scalettar, and Yoshiro Takahashi for fruitful discussions. We particularly thank Tizian Blatz for providing the global subspace expansion code for the time dynamics. This research was funded by the Deutsche Forschungsgemeinschaft (DFG, German Research Foundation) under Germany’s Excellence Strategy—EXC-2111—390814868 and by the European Research Council (ERC) under the European Union’s Horizon 2020 research and innovation programme (grant agreement number 948141 — ERC Starting Grant SimUcQuam). K.R.A.H.'s contributions were supported in part by the Welch Foundation (C-1872), the National Science Foundation (PHY1848304), and the W. M. Keck Foundation (Grant No. 995764).

\appendix
\onecolumngrid
\widetext

\section{The SU(N) t-J model}
\label{sec:tJapp}
We here show that in the case of $N=2$, Eq.~\eqref{eq:H} in the main text is equivalent to the standard $t$-$J$ model without next-nearest neighbor terms, 
\begin{equation}
    \hat{\mathcal{H}} = -t \sum_{\braket{\mathbf{i,j}}, \alpha}  \hat{\mathcal{P}} \left( \hat{c}^{\dagger}_{\alpha, \mathbf{i}} \hat{c}_{\alpha, \mathbf{j}}^{\vphantom\dagger} + \text{h.c.}\right) \hat{\mathcal{P}} + J \sum_{\braket{\mathbf{i,j}}} \left( \hat{S}^x_{\mathbf{i}} \hat{S}^x_{\mathbf{i}} + \hat{S}^y_{\mathbf{i}}\hat{S}^y_{\mathbf{i}} + \hat{S}^z_{\mathbf{i}}\hat{S}^z_{\mathbf{i}} - \frac{1}{4} \hat{n}_{\mathbf{i}} \hat{n}_{\mathbf{j}} \right).
    \label{eq:H_tj}
\end{equation}
Using $\hat{S}^{\mu}_{\mathbf{i}} = \frac{1}{2} \sum_{\alpha \alpha'} \hat{c}^{\dagger}_{\mathbf{i}, \alpha} \sigma_{\alpha \alpha'}^{\mu} \hat{c}_{\mathbf{i}, \alpha'}^{\vphantom\dagger}$ with $\sigma^{\mu}$ ($\mu = x,y,z$) the Pauli matrices, the second part of Eq.~\eqref{eq:H_tj} reads
\begin{equation}
    \begin{aligned}
    \frac{J}{4} \sum_{\braket{\mathbf{i}, \mathbf{j}}} & (\cidup \cidown + \ciddown \ciup )(\cjdup \cjdown + \cjddown \cjup ) + (\cidup \cidown - \ciddown \ciup )(\cjddown \cjup - \cjdup \cjdown ) \\ & + (\cidup \ciup - \ciddown \cidown )(\cjdup \cjup - \cjddown \cjdown ) - (\cidup \ciup + \ciddown \cidown )(\cjdup \cjup + \cjddown \cjdown ),
    \end{aligned}
\end{equation}
which can be rewritten to
\begin{equation}
    \begin{gathered}
    \frac{J}{2} \sum_{\braket{\mathbf{i}, \mathbf{j}}} \cidup \cidown \cjddown \cjup + \ciddown \ciup \cjdup \cjdown + \cidup \ciup \cjdup \cjup + \ciddown \cidown \cjddown \cjdown - (\niup \njdown +\nidown \njup + \niup \njup + \nidown \njdown) \\ = \frac{J}{2} \sum_{\braket{\mathbf{i,j}}} \left(\sum_{\alpha, \beta} 
    \hat{c}^{\dagger}_{\alpha, \mathbf{i}} \hat{c}_{\beta, \mathbf{i}}^{\vphantom\dagger} \hat{c}^{\dagger}_{\beta, \mathbf{j}} \hat{c}_{\alpha, \mathbf{j}}^{\vphantom\dagger} - \hat{n}_{\mathbf{i}} \hat{n}_{\mathbf{j}} \right).
    \end{gathered}
\end{equation}
This corresponds to Eq.~\eqref{eq:H} in the main text. We note that in our calculations, we only implement the hopping and exchange term, i.e., the density-density interaction $-J/2 \sum_{\braket{\mathbf{i}, \mathbf{j}}} \ni \nj$ is not considered. In the case of a single dopant, this only constitutes a constant energy shift in the Hamiltonian in the thermodynamic limit.

\section{Magnetic correlations for varying pinning}
\label{sec:pinningApp}
In the computations in the main text, we have fixed the pinning field to $\mu_p/J = 1$. To test the dependency of the results as the pinning field strength is varied, we calculate on-site flavor occupations $\braket{\hat{n}^f_{\mathbf{i}}}$ in the undoped Heisenberg model for $\mu_p/J = 0.5$ and $\mu_p/J = 1.0$.
\begin{figure*}[h!]
\centering
\includegraphics[width=0.9\textwidth]{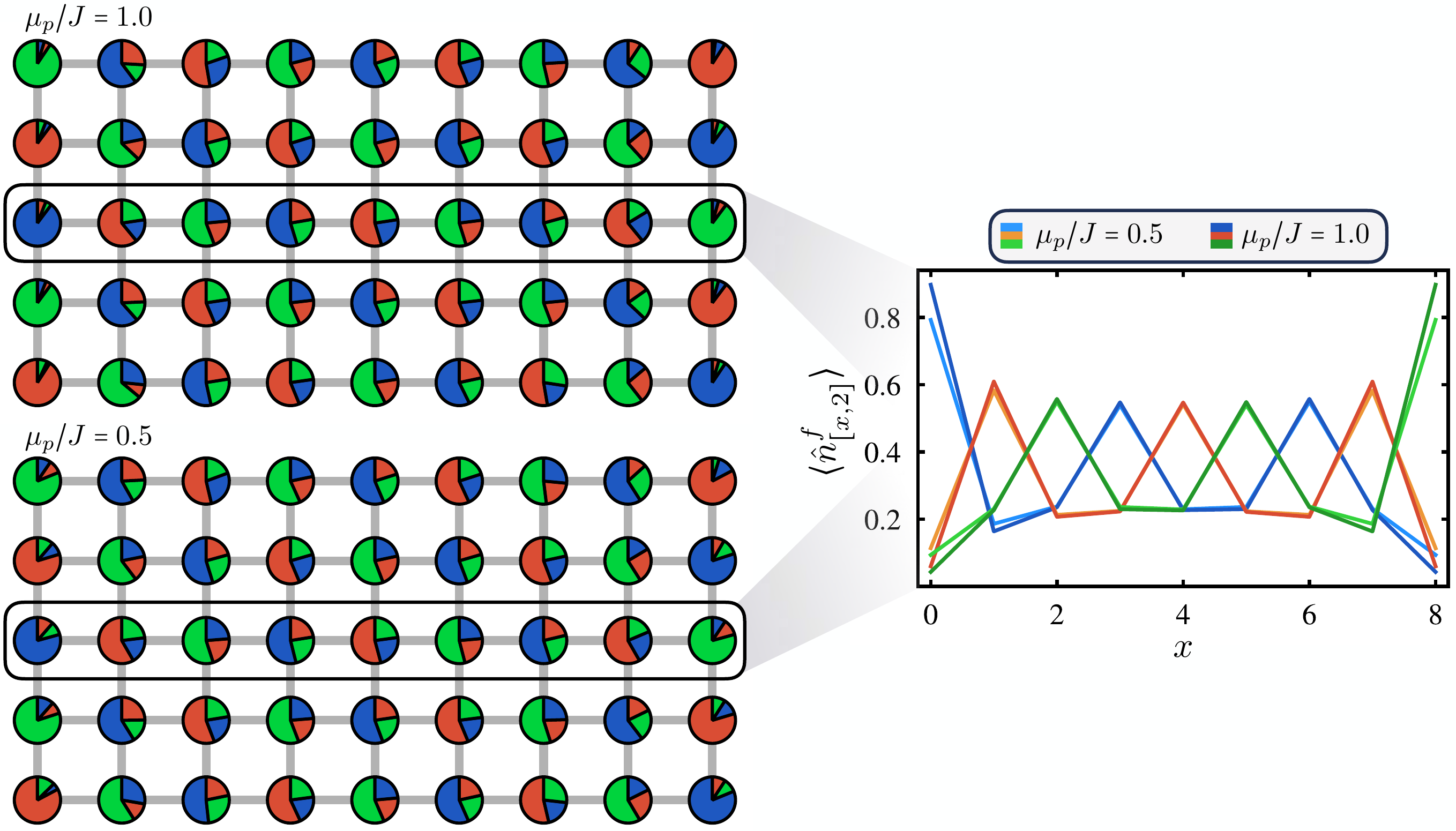}
\caption{\textbf{Varying pinning field.} On-site flavor occupations $\braket{\hat{n}^f_{\mathbf{i}}}$ in the undoped SU(3) symmetric Heisenberg model for $\mu_p/J = 0.5$ and $\mu_p/J = 1.0$. Full occupations are shown on the left-hand side; a cut through the central rung of the system is shown on the right-hand side. While at the boundaries occupations vary, the center of the system is independent on the pinning strength.}
\label{fig:comp_mu}
\end{figure*}
Fig.~\ref{fig:comp_mu} shows the full occupations on the 9x5 lattice (left-hand side) as well as a cut through the central rung of the system (right-hand side). Though directly at the boundaries (where the pinning field is finite) flavor occupations vary as $\mu_p/J$ is changed, in the bulk of the system they are indistinguishable. Hence, we conclude that the long-range order of the magnetic background and hence the physics of the magnetic polaron in the bulk is independent on the pinning strength -- it is merely needed to break the SU(3) symmetry and pin the order.    

\section{Ground state DMRG convergence}
\label{sec:conv_gs}
\begin{figure*}[h!]
\centering
\includegraphics[width=0.8\textwidth]{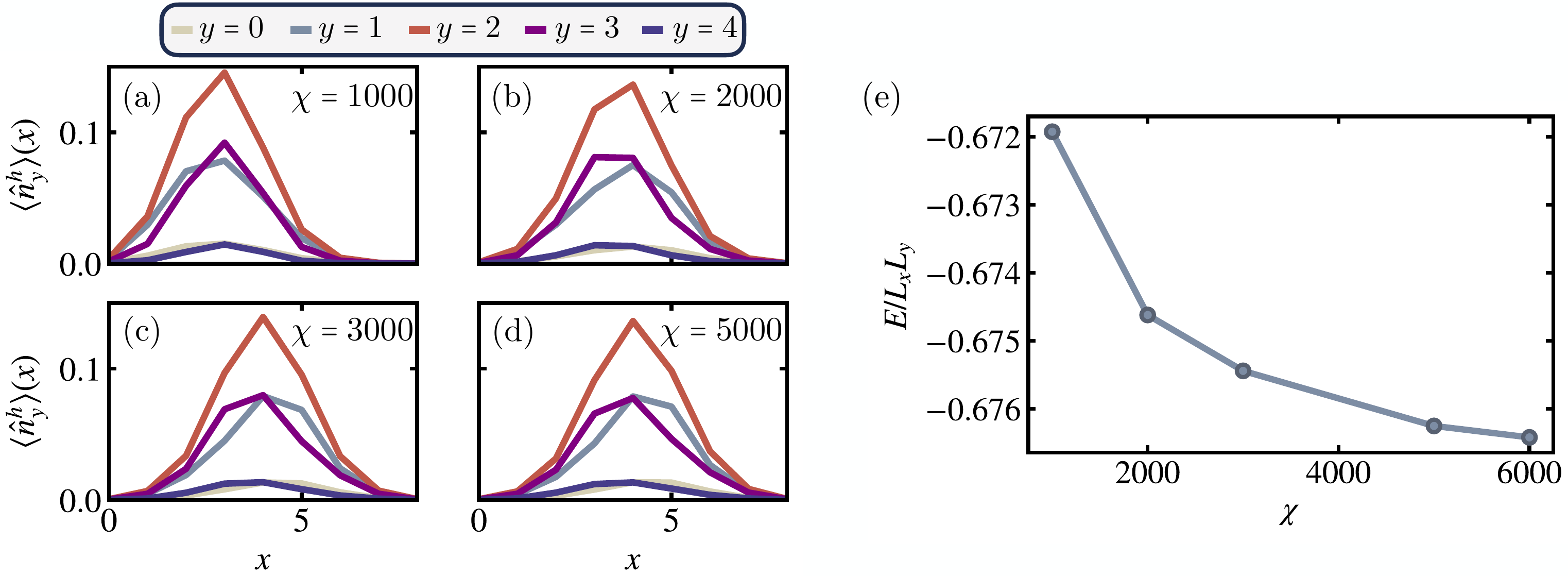}
\caption{\textbf{Ground state DMRG convergence.} Hole densities $n^h_y(x)$ for horizontal cuts $y = 0, \dots,  4$ and bond dimensions $\chi = 1000$ (a), $\chi=2000$ (b), $\chi=3000$ (c) and $\chi=5000$ (d). Convergence is reached for $\chi>3000$. (e) Convergence of the energy per site, E/($L_x L_y$), as a function of bond dimension. Between $\chi = 5000$ and $\chi = 6000$, the relative difference in energies is of the order of a tenth of a percent.}
\label{fig:conv_gs}
\end{figure*}

As mentioned in the main text, we simulate the singly hole-doped SU(3) $t$-$J$ model on a $L_x \times L_y = 9 \times 5$ lattice using open boundary conditions. To break the SU(3) symmetry and pin the 3-SL stripe order, we introduce local chemical potentials at the short edges of the system, $-\mu_p \sum_{i\in \text{edge}} \hat{n}_{\alpha_{\mathbf{i}}, \mathbf{i}}$, with $\mu_p/J = 1$. Fig.~\ref{fig:conv_gs}~(a)-(d) shows the hole densities along horizontal cuts of the system with $t/J = 3$ for increasing bond dimension $\chi = 1000, \dots, 5000$. We observe convergence of the local hole density for bond dimensions $\chi > 3000$, and use $\chi=5000$ throughout the calculations shown in the main text. Similarly, we observe convergence of the energy per site with increasing bond dimension, shown in Fig.~\ref{fig:conv_gs}~(e).  

\section{Dynamics convergence}
\begin{figure*}
\centering
\includegraphics[width=0.38\textwidth]{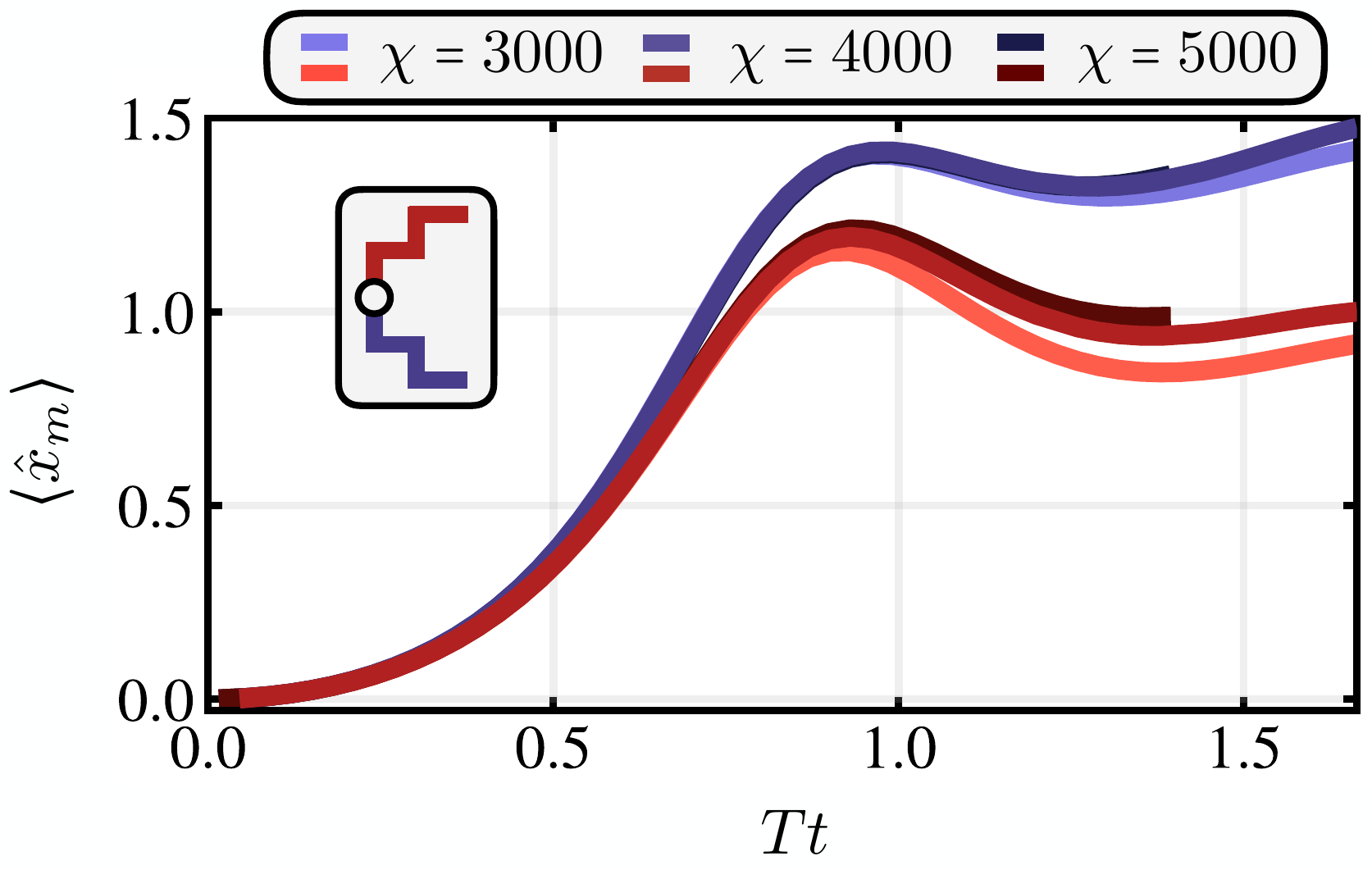}
\caption{\textbf{Convergence of DMRG dynamics.} Mean Manhattan distances along the diagonal and anti-diagonal in a $L_x \times L_y = 9 \times 5$ system with $t/J = 1.5$ as a function of time, for maximum bond dimensions $\chi = 3000,4000,5000$. While the mean Manhattan distance converges along the diagonal, slight differences between $\chi = 4000$ and $\chi = 5000$ are seen along the anti-diagonal. Nevertheless, the dynamical formation of the sub-dimensional magnetic polaron that leads to the appearance of the anisotropy at early times is identical for all bond dimensions.}
\label{fig:conv_dyn}
\end{figure*}

We calculate the dynamical properties of a doped hole using MPS time evolution methods. As local methods suffer from large projection errors at small time steps (where the entanglement in the charge sector is zero), we use the global expansion method~\cite{Yang2020}, before switching to TDVP~\cite{Paeckel_time} once the maximum set bond dimension is achieved. In particular, we choose a Krylov subspace order of 3, time steps $\Delta T J = 0.02$, and bond dimensions $\chi_{\text{max}} = 3000,4000,5000$. Fig.~\ref{fig:conv_dyn} shows the dynamics presented in Fig.~\ref{fig:dynamics} in the main text for the various bond dimensions. Along the diagonal, the mean Manhattan distance is seen to converge up to times $Tt \sim 1.3$. Minor deviations between bond dimensions $\chi = 4000$ and $\chi=5000$ are visible starting from times $Tt \sim 1.0$ along the anti-diagonal. Nevertheless, the early time dynamics including the dynamical appearance of the anisotropy between the diagonal and anti-diagonal mean distances is well converged.

\section{Role of boundary conditions in the undoped $\rm{SU(3)}$ $t$-$J$ model}
\label{sec:BCs}
\begin{figure*}[h!]
\centering
\includegraphics[width=0.8\textwidth]{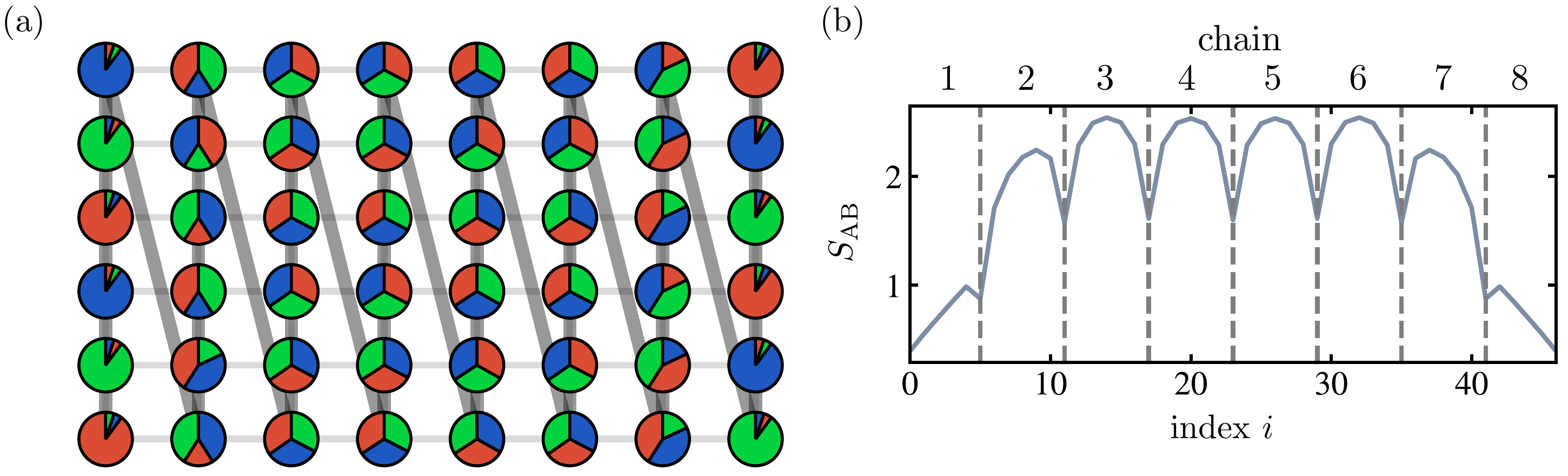}
\caption{\textbf{Periodic boundary conditions.} On-site moments (a) and entanglement entropy (b) for a system of size $L_x \times L_y = 8 \times 6$ with periodic boundaries along the $y$-direction. Both boundaries pin the 3-SL diagonal order, with pinning strength $\mu_p/J = 1$. The site index in (b) indicates the border of the subsystem, following the standard snake indexing as illustrated by dark gray lines in (a). Both entanglement entropy and on-site moments suggest the absence of magnetic order despite the applied pinning, and instead suggest the appearance of coupled 1D (periodic) chains.}
\label{fig:PBC}
\end{figure*}

As mentioned in the main text, we focus on open boundary conditions (OBC) in both directions in our simulations. Indeed, we find that for the accessible system sizes OBCs are crucial to observe the three sublattice diagonal stripe order in the undoped ground state of the $\rm{SU(3)}$ $t$-$J$ model, consistent with what was mentioned in Ref.~\cite{Bauer2012}. 
Fig.~\ref{fig:PBC}~(a) shows the on-site moments of the three flavors for a system of size $L_x \times L_y = 8, 6$ with periodic boundaries (PBC) applied along the short (y-) direction. Even with the applied pinning (shown in Fig.~\ref{fig:PBC} for $\mu_p/J = 1$), the order rapidly disappears away from the boundaries. 
Moreover, we have carefully checked that full projections of real-space patterns do further not reveal any ordered state. For the system widths considered here, the entanglement entropy reveals that the ground state converges to a state of weakly coupled 1D (periodic) chains, as shown in Fig.~\ref{fig:PBC}~(b). We conclude that the appearance of weakly coupled chains in periodic systems is an artifact of finite-size effects, and that we expect diagonal stripe order to appear when $L_y \gg 1$ becomes much larger -- which is, however, not accessible with current numerical techniques. 

\twocolumngrid
\bibliography{SU3}

\end{document}